\documentclass[twocolumn,showpacs,preprintnumbers,amsmath,amssymb,floatfix,a4paper]{revtex4}

\usepackage{graphicx}
\usepackage{dcolumn}
\usepackage{bm}
\usepackage{setspace}
\usepackage{amsmath}

\begin{document}

\title{A revised thermonuclear rate of $^{7}$Be($n$,$\alpha$)$^{4}$He relevant to Big-Bang nucleosynthesis}

\author{S.Q.~Hou$^{1,2}$}
\author{J.J.~He$^1$}
\email{jianjunhe@impcas.ac.cn}
\author{S.~Kubono$^{1,3}$}
\author{Y.S.~Chen$^{4}$}

\affiliation{$^1$Key Laboratory of High Precision Nuclear Spectroscopy and Center for Nuclear Matter Science, Institute of Modern Physics, Chinese Academy of Sciences, Lanzhou 730000, China}
\affiliation{$^2$University of Chinese Academy of Sciences, Beijing 100049, China}
\affiliation{$^3$RIKEN Nishina Center, 2-1 Hirosawa, Wako, Saitama 351-0198, Japan}
\affiliation{$^4$China Institute of Atomic Energy, P. O. Box 275(10), Beijing 102413, China}

\date{\today}

\begin{abstract}
In the standard Big-Bang nucleosynthesis (BBN) model, the primordial $^7$Li abundance is overestimated by about a factor of 2--3 comparing to the astronomical
observations, so called the pending cosmological lithium problem. The $^7$Be($n$,$\alpha$)$^4$He reaction, which may affect the $^7$Li abundance, was regarded
as the secondary important reaction in destructing the $^7$Be nucleus in BBN. However, the thermonuclear rate of $^7$Be($n$,$\alpha$)$^4$He has not been well
studied so far. This reaction rate was firstly estimated by Wagoner in 1969, which has been generally adopted in the current BBN simulations and the reaction
rate library. This simple estimation involved only a direct-capture reaction mechanism, but the resonant contribution should be also considered according to the
later experimental results. In this work, we have revised this rate based on the indirect cross-section data available for the $^4$He($\alpha$,$n$)$^7$Be and
$^4$He($\alpha$,$p$)$^7$Li reactions, with the charge symmetry and detailed-balance principle. Our new result shows that the previous rate (acting as an upper
limit) is overestimated by about a factor of ten. The BBN simulation shows that the present rate leads to a 1.2\% increase in the final $^7$Li abundance compared
to the result using the Wagoner rate, and hence the present rate even worsens the $^7$Li problem. By the present estimation, the role of
$^7$Be($n$,$\alpha$)$^4$He in destroying $^7$Be is weakened from the secondary importance to the third, and the $^7$Be($d$,$p$)2$^4$He reaction becomes of
secondary importance in destructing $^7$Be.
\end{abstract}


\maketitle

\section{Introduction}
\label{sec:introduction}

The discrepancy between the predicted primordial $^7$Li abundance and the astronomical observation persists as a fundamental pending problem in nuclear
astrophysics studies~\cite{Cyburt03,Coc04,Fields12}. As a powerful tool to study the early universe, the standard Big-Bang nucleosynthesis (BBN) model employs
only one parameter $\eta$, the number-density ratio of baryons to photons. With the more accurate $\eta$ value determined from the astronomical
observations~\cite{WMAP5}, the predicted primordial $^7$Li abundance is still a factor of 2--3 higher than that observed in the galactic halo
stars~\cite{Cyburt08}. It has been argued that such discrepancy may arise from the uncertainties in the thermonuclear rates for those reactions involved in
BBN~\cite{Coc04,Nollett00,Fields12,Cha11}. In the past two decades, great efforts have been devoted to reduce these uncertainties. For example, Smith et
al.~\cite{Micheal93} made a new evaluation of the reaction rates for the most important twelve reactions involved in BBN. Later on, Descouvemout
et al.~\cite{Disc04} re-analyzed ten key reactions by using the $R$-matrix theory. However, the $^7$Li discrepancy still remains unsolved with these updated
data together with the recent investigations~\cite{Angulo05,Cha11,Hamma,Jcap12} for those possible reactions affecting the $^7$Li abundance.

The final $^7$Li abundance in BBN is contributed both from the directly synthesized $^7$Li, as well as those from the $^7$Be EC decays and $^7$Be($n$,$p$)$^7$Li
reaction, however, the relic $^7$Li nuclei mainly come from the latter process because most of the directly synthesized $^7$Li is destroyed through the
$^7$Li($p$,$\alpha$)$^4$He reaction immediately. The $^7$Be production is determined by the balance between the reactions which synthesize it and those destruct
it. Therefore, it is essential to accurately determine thermonuclear rates for those reactions involving $^7$Be nuclide. The key synthesizing reaction for $^7$Be
is the $^3$He($\alpha$,$\gamma$)$^7$Be reaction, which has been studied very well by various experiments and theories~\cite{CyburtBe7,Kon13,Neff11,Luna07}.
The reactions for destructing $^7$Be are $^7$Be($n$,$p$)$^7$Li and $^7$Be($n$,$\alpha$)$^4$He, of primary and secondary importance, respectively. For the
$^7$Be($n$,$p$)$^7$Li reaction, its cross section has been studied in detailed in a wide energy range from 0.025 eV up to 8 MeV~\cite{Ada03,Disc04}, which covers
entirely the BBN effective energy region. As for the $^7$Be($n$,$\alpha$)$^4$He reaction, it could play a non-negligible role in direct $^7$Be
destruction~\cite{Serpico04}. However, up to now, the $^7$Be($n$,$\alpha$)$^4$He reaction rate adopted in the current BBN simulations and the reaction rate
library~\cite{JINA} is still the very old Wagoner rate~\cite{Wag69}, which is a simple theoretical estimation involving only the direct-capture reaction
mechanism, and also without information on the sources of data and error estimate. In this work, we have derived the thermonuclear rate of
$^7$Be($n$,$\alpha$)$^4$He based on the available indirect experimental data of $^4$He($\alpha$,$n$)$^7$Be and $^4$He($\alpha$,$p$)$^7$Li reactions, with the
well-know charge symmetry and detailed-balance principle~\cite{Book52}. With this new rate, we have examined its impact on the primordial $^7$Li abundance with
a BBN code.

\section{Derivation of $^7$Be($n$,$\alpha$)$^4$He cross section}
By so far, there is only one direct cross-section measurement for the $^7$Be($n$,$\alpha$)$^4$He reaction. The experiment was performed by Bassi
et al.~\cite{Bassi63} in 1963 by using the reactor thermal neutrons. Based on this limited information and the theory of nonresonant reaction, Wagoner made the
first estimation of this reaction rate, which will be discussed in detail in the next section. In this section, we will present the method to derive the cross
section of $^7$Be($n$,$\alpha$)$^4$He by using the available indirect experimental data.

About 30 years ago, King et al.~\cite{King77} measured both cross sections of $^4$He($\alpha$,$n$)$^7$Be and $^4$He($\alpha$,$p$)$^7$Li. Under the assumption
of charge symmetry, i.e., the neutron and proton configurations in the compound $^8$Be nuclide to be identical, they calculated the total cross sections of
$^4$He($\alpha$,$n$)$^7$Be based on their data measured for $^4$He($\alpha$,$p$)$^7$Li via the following equation,
\begin{equation}
\sigma_n=\sigma_{n_0} +\sigma_{n_1}= \frac{P^{n_0}_{\ell}}{P^{p_0}_{\ell}}\sigma_{p_0} + \frac{P^{n_1}_{\ell}}{P^{p_1}_{\ell}}\sigma_{p_1},
\label{eq:one}
\end{equation}
where $\sigma_{n_0}$ and $\sigma_{p_0}$ are the cross sections leading to the ground states of $^7$Be and $^7$Li, respectively; and $\sigma_{n_1}$ and
$\sigma_{p_1}$ are those leading to the corresponding first excited states. Here, $P_\ell$ is the penetrability factor defined by~\cite{King77,Illidis07}:
\begin{eqnarray}
P_\ell(E,R)=\frac{kR}{F^2_\ell(E,R)+G^2_\ell(E,R)},
\label{eq:two}
\end{eqnarray}
where $k$ is the wave number, $R$ the channel radius, and $F_\ell$ and $G_\ell$ the standard Coulomb functions. It was shown that the calculated $\sigma_n$
agreed with their experimental data very well.

Inspired by this idea, we have derived the cross section of $^4$He($\alpha$,$n$)$^7$Be based on those measured for $^4$He($\alpha$,$p$)$^7$Li~\cite{King77,Slob82},
as well as those measured for $^7$Li($p$,$\alpha$)$^4$He~\cite{CA63} with the detailed-balance principle.
The experimental cross-section data of $^4$He($\alpha$,$p$)$^7$Li and those derived from $^7$Li($p$,$\alpha$)$^4$He are listed in the first and second columns
of Table~\ref{tab1}.

\begin{table}
\caption{\label{tab1} Experimental cross-section data collected for $^4$He($\alpha$,$p$)$^7$Li (g.s.), and the corresponding ones derived for
$^7$Be($n$,$\alpha$)$^4$He, in units of mb. The adopted uncertainties in energies (in units of MeV) of $E_{\alpha}$ and $E_\mathrm{c.m.}$ are $\pm$100 keV and
$\pm$50 keV, respectively~\cite{King77,Slob82}.}
\begin{ruledtabular}
\begin{tabular}{rrrrr}
$E_{\alpha}$       &  $\sigma_{(\alpha,p)}$   & $E_\mathrm{c.m.}$  &  $\sigma_{(n,\alpha)}$   &  Ref.  \\
\hline
      &               & 0.0113       &    8.4$\pm$8.5  &  \cite{DirectLi7p}     \\
      &               & 0.0196       &   10.7$\pm$10.7 &  \cite{DirectLi7p}     \\
      &               & 0.0510       &   13.2$\pm$13.2 &  \cite{DirectLi7p}¡¡   \\
38.23 & 13.0$\pm$0.4\footnotemark[1] & 0.124  &   17.5$\pm$10.4  & \cite{CA63}  \\
38.41 & 14.9$\pm$1.4\footnotemark[1] & 0.214  &   23.1$\pm$8.2   & \cite{CA63}  \\
38.54 & 24.2$\pm$2.0  & 0.279 &   39.0$\pm$11.1  &\cite{Slob82}    \\
38.96 & 35.5$\pm$2.5  & 0.489 &   59.4$\pm$10.3  &\cite{Slob82}    \\
38.97 & 29.9$\pm$1.5  & 0.494 &   50.0$\pm$8.2   &\cite{King77}    \\
39.44 & 49.2$\pm$3.1  & 0.729 &   79.1$\pm$9.4   & \cite{Slob82}   \\
39.80 & 59.9$\pm$3.0  & 0.909 &   91.6$\pm$8.3   &\cite{King77}    \\
39.94 & 64.6$\pm$2.6  & 0.979 &   96.8$\pm$7.6   &\cite{Slob82}    \\
40.56 & 30.5$\pm$2.5  & 1.289 &   41.5$\pm$3.9   &\cite{Slob82}    \\
40.99 & 27.0$\pm$2.2  & 1.504 &   34.4$\pm$3.0   &\cite{Slob82}    \\
41.35 & 23.4$\pm$1.2  & 1.684 &   28.2$\pm$1.6   &\cite{King77}    \\
41.61 & 17.9$\pm$2.0  & 1.814 &   20.8$\pm$2.4   &\cite{Slob82}    \\
41.95 & 12.0$\pm$0.6  & 1.984 &   13.3$\pm$0.7   &\cite{King77}    \\
42.57 & 6.5 $\pm$0.3  & 2.294 &    6.6$\pm$0.3   &\cite{King77}    \\
43.04 & 13.1$\pm$2.1  & 2.529 &   12.6$\pm$2.0   &\cite{Slob82}    \\
43.52 & 12.0$\pm$0.6  & 2.769 &   11.0$\pm$0.6   &\cite{King77}    \\
44.32 & 52.0$\pm$2.6  & 3.169 &   43.6$\pm$2.2   &\cite{King77}    \\
45.64 & 36.5$\pm$1.8  & 3.829 &   27.1$\pm$1.4   &\cite{King77}    \\
46.67 & 27.2$\pm$1.4  & 4.344 &   18.6$\pm$1.0   &\cite{King77}    \\
47.65 & 22.7$\pm$1.1  & 4.884 &   14.3$\pm$0.7   &\cite{King77}    \\
49.49 & 15.1$\pm$0.8  & 5.754 &    8.6$\pm$0.5   &\cite{King77}    \\
\end{tabular}
\end{ruledtabular}
\footnotetext[1]{The cross-section data are derived from those of $^7$Li($p$,$\alpha$)$^4$He with Eq.~\ref{eq:one}. Since their incident energies have no errors
quoted in the original paper~\cite{CA63}, and we assume the same errors as in Refs.~\cite{King77,Slob82}.}
\end{table}

In the astrophysical high-temperature environment the excited states of nuclei involved are thermally populated, which can also contribute to the total reaction
rate~\cite{Bahcall69,Rolfs88}. The first excited state in $^7$Be is located at 429 keV, which is too high to make appreciable contribution in the total rate
comparing to the ground state at BBN temperature region. Thus, it is appropriate to calculate the thermonuclear rate of $^7$Be(n,$\alpha$)$^4$He by taking only
the ground state of $^7$Be nucleus into account. Relying on the first term of Eq.~\eqref{eq:one}, we have directly derived the cross section of
$^4$He($\alpha$,$n$)$^7$Be (g.s.) by utilizing those data of $^4$He($\alpha$,$p$)$^7$Li (g.s.) listed in Table~\ref{tab1}.
It is worthy of noting that the ground-state contribution required is difficult to be extracted from the total cross section measured~\cite{King77} for
$^4$He($\alpha$,$n$)$^7$Be, and also these data were only measured down to $E_\alpha$=39.43 MeV which is much higher than those measured for
$^4$He($\alpha$,$p$)$^7$Li.

In Eq.~\eqref{eq:two}, the penetrability factor depends on the channel radius $R$, orbit angular momentum $\ell$ and incident center-of-mass energy $E$.
In the treatment of King et al., a channel radius of $R$=4.1 fm [i.e., $R$=$r_0$(A$_1^{1/3}$+A$_2^{1/3}$) with $r_0$=1.41 fm] was utilized in both
$n$+$^7$Be and $p$+$^7$Li systems. In the present calculation, the penetrability factor $P_\ell(E,R)$ for $p$+$^7$Li is calculated by an RCWFN
code~\cite{RCWFN}, and that for neutron is calculated using Eq.~\eqref{eq:two} with $F_\ell$ and $G_\ell$ values tubulated by Feshbach and Lax~\cite{Lax48}.
Since the $\alpha$ particle is the spinless boson, the wavefunction for two identical $\alpha$ particles must be symmetric under interchange. However, the
wavefunction for an $\ell$=\textit{odd} state in $^8$Be is antisymmetric by interchanging the two $\alpha$ particles. This implies that the compound state in
$^8$Be must have even parity for the incident $\alpha$+$\alpha$ channel. We know that both ground states for $^7$Li and $^7$Be have odd parity, and hence the
relative orbit angular momentum $\ell$ must be \textit{odd}. Since the orbit centrifugal barrier [$\propto$$\ell$($\ell$+1)] increases with respect to $\ell$,
the $p$-wave ($\ell$=1) capture will dominate both the $n$+$^7$Be and $p$+$^7$Li exit channels. For each energy points listed in Table~\ref{tab1}, we have
calculated the $^4$He($\alpha$,$n$)$^7$Be (g.s.) reaction cross section, and the associated uncertainty is estimated by considering the uncertainties in both
$r_0$ (in range of 1.1--1.5 fm~\cite{r0}) and incident energy ($\pm$100 keV~\cite{King77,Slob82}). Finally, with the detailed-balance principle~\cite{Book52}
the cross section of $^7$Be($n$,$\alpha$)$^4$He has been derived as listed in the third and fourth columns of Table~\ref{tab1}. Here, three data points are
listed for $E_\mathrm{c.m.}$$<$0.1 MeV, which are derived based on the experimental data~\cite{DirectLi7p} of $^7$Li($p$,$\alpha$)$^4$He simply by the first
term in Eq.~\ref{eq:one}. The associated uncertainties are estimated by taking those of $r_0$ and incident energies into account. Alternatively, the low-energy
data of $^4$He($\alpha$,$n$)$^7$Be have been simply estimated by linearly interpolating two data points of $^4$He($\alpha$,$p$)$^7$Li at 37.48 MeV~\cite{Slob82}
and 38.09 MeV~\cite{CA63}, and ultimately converted to those of $^7$Be($n$,$\alpha$)$^4$He. And we find that the interpolated results agree well with these three
data points at $E_\mathrm{c.m.}$$<$0.1 MeV within uncertainties.

\section{Wagoner's estimation}
\begin{figure}
\resizebox{8.3cm}{!}{\includegraphics{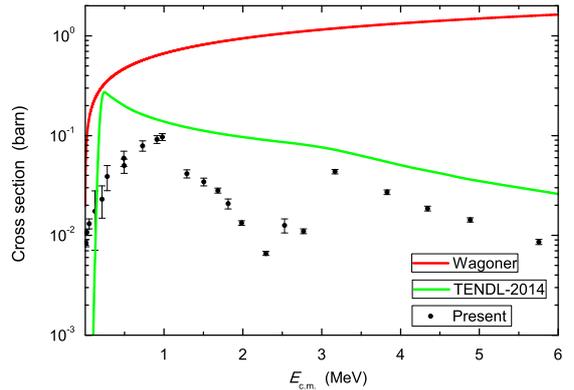}}
\caption{\label{fig1} (Color online) Comparison of different cross sections of $^7$Be(n,$\alpha$)$^4$He among present work and other two different
origins (Wagoner's work~\cite{Wag69} and TENDL-2014 evaluation~\cite{TENDL}).}
\end{figure}

The thermonuclear rate of $^7$Be($n$,$\alpha$)$^4$He has only been estimated by Wagoner~\cite{Wag69} in 1969, which has been generally adopted in the current
BBN simulations and the reaction rate library~\cite{JINA}. In Wagoner's paper, the reaction rate was calculated by an analytical formula
\begin{equation}
\label{eq:thr}
N_A\left\langle\sigma v\right\rangle=2.05\times10^4\times(1+3760T_9).
\end{equation}
For the non-resonant neutron-induced reaction, $<$$\sigma v$$>$ can be expressed by~\cite{FCZ67,Rolfs88,Illidis07}
\begin{equation}
\label{eq:five}
\left\langle\sigma v\right\rangle=\mathcal{S}(0)+0.3312\dot{\mathcal{S}}(0)T_9^{1/2}+0.06463\ddot{\mathcal{S}}(0)T_9,
\end{equation}
with $\mathcal{S}$(0) the astrophysical $S$ factor near zero energy. By comparing Eq.~\ref{eq:thr} and Eq.~\ref{eq:five}, the values of $\mathcal{S}$(0) and
and it's second derivative $\ddot{\mathcal{S}}$(0) (\textit{wrt} velocity $v$) can be obtained by assuming a negligible first derivative $\dot{\mathcal{S}}$(0).
Thus, the direct-capture cross section for neutron-induced reaction can be calculated by~\cite{FCZ67,Illidis07},
\begin{equation}
\label{eq:three}
\sigma=\frac{\mathcal{S}(0)}{v}=\frac{\mathcal{S}(0)+\frac{1}{2}\ddot{\mathcal{S}}(0)E}{v}.
\end{equation}
In this way, the cross section of $^7$Be($n$,$\alpha$)$^4$He is calculated as shown in Fig.~\ref{fig1}.

There is no any information on the sources of $^7$Be($n$,$\alpha$)$^4$He cross-section data used in the Wagoner's original paper~\cite{Wag69}. Before his
estimation, there was only one measurement made by Bassi et al.~\cite{Bassi63} at thermal neutron energy. Here, we have figured out the Wagoner's way of thinking.
The total cross section of $^7$Be($n$,$\alpha$)$^4$He was assumed to be composed of two contributions, i.e., $\sigma_1$+$\sigma_2$, with $\sigma_1$ for
the $p$-wave capture of ($n$,$\alpha$) and $\sigma_2$ for the $s$-wave capture of ($n$,$\gamma\alpha$), respectively. The former obeys the law of
$\sigma_2$$\propto$$v$, and the latter obeys the law of $\sigma_1$$\propto$$1/v$~\cite{Illidis07,Vogl64}. According to this way, Wagoner made his estimation
based on an upper limit of $\sigma_1$$\leq$0.1 mb and $\sigma_2$=155 mb measured by Bassi et al. at thermal neutron energy. Actually, the $p$-wave dominates the
cross section (or total rate) at the energy region of BBN relevant. Therefore, Wagoner actually gave us only an upper limit.

The Wagoner cross sections and ours are compared in Fig.~\ref{fig1}. It shows that the present results are overall lower than those of Wagoner (upper limit).
However, our results are only derived based on the indirect experimental cross-section data (including both the direct- and resonant-capture contributions),
further measurements~\cite{gale,cern} are proposed to acquire the direct experimental data.
In addition, the recent theoretical evaluation of TENDL-2014~\cite{TENDL} based on a TALYS calculation is also compared, which is entirely different from the
present results as shown in Fig.~\ref{fig1}.

\section{Revised thermonuclear rate}
The thermonuclear $^7$Be($n$,$\alpha$)$^4$He rate as a function of temperature was calculated by numerical integration of our experimental cross sections
using the EXP2RATE code by T.~Rauscher~\cite{Exp2}. Rate values, obtained as the arithmetic mean between the low and high limits associated with the
uncertainties on both the cross-section data and incident energies, are given in Table~\ref{tab2}. The present rate can be well parameterized (less than 0.4\%
error in 0.1--5 GK) by the following expression (e.g., in the standard format of Eq.~(16) in Ref.~\cite{rau00}):

\begin{widetext}
\begin{eqnarray}
N_A\langle\sigma v\rangle=\mathrm{exp}(-17.8984+0.2711T_9^{-1}-23.8918T_9^{-1/3}+62.2135T_9^{1/3}-5.2888T_9+0.3869T_9^{5/3}-22.6197\ln{T_9}).
\label{eq:fit}
\end{eqnarray}
\end{widetext}

Our new rate is about a factor of ten smaller than the Wagoner rate in BBN temperature range. As discussed above, Wagoner just presented an upper limit for this
rate. Therefore, we propose here that the cross section $\sigma_1$ of $^7$Be($n$,$\alpha$)$^4$He at the thermal neutron energy is about 0.01 mb, one order of
magnitude smaller than the previous upper limit~\cite{Bassi63}.

\begin{table}
\caption{\label{tab2} Thermonuclear reaction rates for $^7$Be($n$,$\alpha$)$^4$He in units of cm$^3$s$^{-1}$mol$^{-1}$. The ratio between present rate and
Wagoner rate is listed in the last column.}
\begin{ruledtabular}
\begin{tabular}{cccc}
$T$(GK)      & Present     & Wagoner    & Ratio \\
\hline
0.1 &	(9.6$\pm$8.3)$\times$10$^5$  &	7.7$\times$10$^6$  &   0.13  \\
0.2 &	(1.7$\pm$1.3)$\times$10$^6$  &	1.5$\times$10$^7$  &   0.11  \\
0.3 &	(2.3$\pm$1.7)$\times$10$^6$  &	2.3$\times$10$^7$  &   0.10  \\
0.4 &	(2.9$\pm$2.0)$\times$10$^6$  &	3.1$\times$10$^7$  &   0.09  \\
0.5 &	(3.5$\pm$2.2)$\times$10$^6$  &	3.9$\times$10$^7$  &   0.09  \\
0.6 &	(4.2$\pm$2.4)$\times$10$^6$  &	4.6$\times$10$^7$  &   0.09  \\
0.7 &	(4.9$\pm$2.6)$\times$10$^6$  &	5.4$\times$10$^7$  &   0.09  \\
0.8 &	(5.6$\pm$2.8)$\times$10$^6$  &	6.2$\times$10$^7$  &   0.09  \\
0.9 &	(6.4$\pm$2.9)$\times$10$^6$  &	6.9$\times$10$^7$  &   0.09  \\
1.0 &	(7.2$\pm$3.1)$\times$10$^6$  &	7.7$\times$10$^7$  &   0.09  \\
1.5 &	(1.2$\pm$0.7)$\times$10$^7$  &	1.2$\times$10$^8$  &   0.10  \\
2.0 &	(1.7$\pm$0.4)$\times$10$^7$  &	1.5$\times$10$^8$  &   0.11  \\
2.5 &	(2.1$\pm$0.4)$\times$10$^7$  &	1.9$\times$10$^8$  &   0.11  \\
3.0 &	(2.5$\pm$0.5)$\times$10$^7$  &	2.3$\times$10$^8$  &   0.11  \\
3.5 &	(2.9$\pm$0.5)$\times$10$^7$  &	2.7$\times$10$^8$  &   0.11  \\
4.0 &	(3.2$\pm$0.5)$\times$10$^7$  &	3.1$\times$10$^8$  &   0.10  \\
4.5 &	(3.4$\pm$0.5)$\times$10$^7$  &	3.5$\times$10$^8$  &   0.10  \\
5.0 &	(3.5$\pm$0.5)$\times$10$^7$  &	3.9$\times$10$^8$  &   0.09  \\
\end{tabular}
\end{ruledtabular}
\end{table}

\begin{figure*}[t]
\begin{center}
\includegraphics[width=10cm]{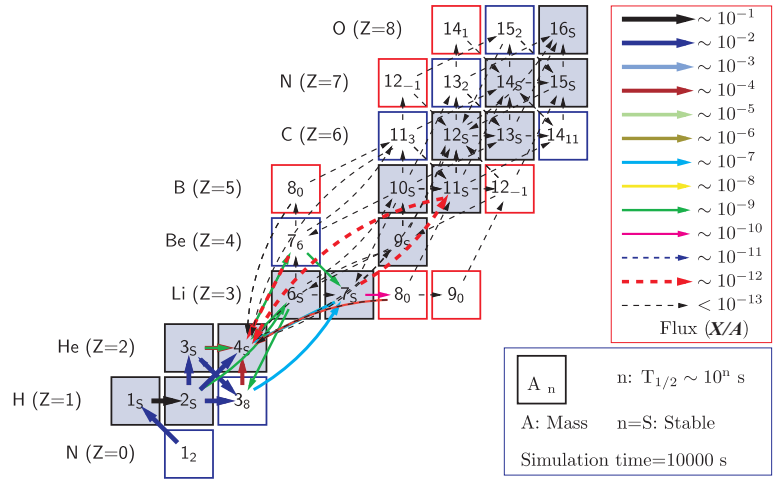}
\vspace{-3mm}
\caption{\label{fig2} (Color online) The time-integrated reaction flow for BBN network with new rate of $^7$Be($n$,$\alpha$)$^4$He. Here, $X$ and $A$ denote the
mass fraction and molar mass, respectively.}
\end{center}
\end{figure*}

\section{BBN simulation}
We have investigated the impact of our new rates on the BBN predicted abundances of D, $^3$He, $^4$He and $^7$Li by using a recently developed code~\cite{Hou10}.
This code can calculate the reaction flux for every specific reaction at arbitrary time point. In this work, the recent values for cosmological parameters and
nuclear physics quantities, such as the baryon-to-photon ratio $\eta$=(6.203$\pm$0.137)$\times$10$^{-10}$~\cite{WMAP9} and the neutron life-time
$\tau$=880.3 s~\cite{Tau14}, have been utilized in our model. The number of light neutrino families N$_{\nu}$=2.9840$\pm$0.0082 determined by CERN LEP
experiment\cite{NvKind} supports the standard model prediction of $N_{\nu}$=3, which is adopted in the present calculation. The reaction network has the same
size as in Ref.~\cite{Hou10} with nuclei $A\le$16 from $n$ to $^{16}$O, and the relevant reaction rates are adopted from the
literature~\cite{Nacre99,Disc04,Micheal93,Wag69,CyburtBe7,Serpico04}.

Here, two simulations have been performed with the Wagoner rate and our new rate of the $^7$Be($n$,$\alpha$)$^4$He reaction studied, respectively. It shows that
the predicted abundances of D, $^3$He and $^4$He do not change appreciably, while that for $^7$Li increases 1.2\% by adopting the new rate. Therefore, the
present rate even worsens the $^7$Li problem. In order to clarify the reason, we have performed the reaction flux~\cite{Illidis07} calculations. Figure~\ref{fig2}
shows the flow passing though arbitrary reaction involved in the BBN network in a timescale of about 10000 s (long enough for BBN) with the present rate. The
flux passing through the $^7$Be($n$,$\alpha$)$^4$He channel is about 10$^{-12}$ mol/g (marked by a dashed red arrow), which is about a factor of ten weaker than
the result using the Wagoner rate. Based on the present estimation, the role of $^7$Be($n$,$\alpha$)$^4$He in destroying $^7$Be is weakened from the secondary
importance to the third, and thus the $^7$Be($d$,$p$)2$^4$He reaction becomes the secondary important reaction in destructing $^7$Be.
In addition, our calculation shows that the cosmological $^7$Li problem could be solved (within the observation uncertainties) provided only if that the
thermonuclear rate of $^7$Be($n$,$\alpha$)$^4$He is about 180 times larger than the Wagoner rate under the condition of without changing the rates for those
rest reactions. Therefore, it is unlikely to solve the $^7$Li problem by a further study of this reaction.

\section{Conclusion}
We have revised the thermonuclear rate of $^7$Be($n$,$\alpha$)$^4$He, which was regarded as the secondary important reaction in destructing the $^7$Be nucleus in
BBN, and ultimately tried to understand the cosmological $^7$Li problem better. The present work shows that the previous Wagoner's result is overestimated
by about a factor of ten. The BBN simulation shows that the adoption of the new rate can not yield appreciable change to the final $^7$Li abundance, only about
1.2\% enhancement, which even worsens the $^7$Li problem. The resolution for this mysterious problem might resort to other mechanisms or new physics beyond the
standard models (e.g., see Refs.~\cite{kusa13,coc13}). Another possibility is that the current observational data might not exactly represent the primordial
$^7$Li abundance. The detailed discussion about how to solve this pending problem is beyond the scope of this paper.

\begin{center}
\textbf{Acknowledgments}
\end{center}
This work was financially supported by the National Natural Science Foundation of China (Nos. 11490562, 11135005, 11321064) and the Major State Basic Research
Development Program of China (2013CB834406). S.K acknowledges support from the Visiting Professorships Program of CAS (No. 2012T1J0013) and JSPS KAKENHI
(No. 26287058) in Japan. S.Q.H would like to express the appreciation to K.M. Nollett (University of South Carolina, Columbia) for meaningful discussions.


\begin{thebibliography}{99}
\bibitem{Cyburt03}
R.H. Cyburt {\it et al.}, Phys. Lett. B \textbf{567}, 227 (2003).
\bibitem{Coc04}
A. Coc {\it et al.}, Astrophys. J. \textbf{600}, 544 (2004).
\bibitem{Fields12}
B.D. Fields, Annu. Rev. Nucl. Part. Sci. \textbf{61}, 47 (2011).
\bibitem{WMAP5}
J. Dunkley {\it et al.},  Astrophys. J. Suppl. \textbf{180}, 306 (2009).
\bibitem{Cyburt08}
R.H. Cyburt, B.D. Fields and K.A. Olive, J. Cosmol. Astropart. Phys. \textbf{11}, 012 (2008).
\bibitem{Nollett00}
K.M. Nollett and S. Burles, Phys. Rev. D, \textbf{61}, 123505 (2000).
\bibitem{Cha11}
N. Chakraborty, B.D. Fields and K.A. Olive, Phys. Rev. D \textbf{83}, 063006 (2011).
\bibitem{Micheal93}
M.S. Smith, L.H. Kawano and R.A. Malaney, Astrophys. J. Suppl. \textbf{85}, 219 (1993).
\bibitem{Disc04}
P. Descouvemont {\it et al.}, At. Data Nucl. Data Tables \textbf{88}, 203 (2004).
\bibitem{Hamma}
F. Hammache {\it et al.}, Phys. Rev. C \textbf{88}, 062802(R) (2013).
\bibitem{Jcap12}
C. Broggini {\it et al.}, J. Cosmol. Astropart. Phys. \textbf{06}, 030 (2012).
\bibitem{Angulo05}
C. Angulo {\it et al.}, Astrophys. J. \textbf{630}, L105 (2005).
\bibitem{Luna07}
Gy. Gyurky {\it et al.}, Phys. Rev. C \textbf{75}, 035805 (2007).
\bibitem{CyburtBe7}
R.H. Cyburt and B. Davids, Phys. Rev. C \textbf{78}, 064614 (2008).
\bibitem{Neff11}
T. Neff, Phys. Rev. Lett. \textbf{106}, 042502 (2011).
\bibitem{Kon13}
A. Kontos {\it et al.}, Phys. Rev. C \textbf{87}, 065804 (2013).
\bibitem{Ada03}
A. Adahchour and P. Descouvemont, J. Phys. G \textbf{29}, 395 (2003).
\bibitem{Serpico04}
P.D. Serpico {\it et al.}, J. Cosmol. Astropart. Phys. \textbf{12}, 010 (2004).
\bibitem{JINA}
JINA REACLIB Database, https://groups.nscl.msu.edu/jina/reaclib/db/.
\bibitem{Wag69}
R.V. Wagoner, Astrophys. J. Suppl. \textbf{18}, 247 (1969).
\bibitem{Book52}
J.M. Blatt and V.F. Weisskopf, Theoretical nuclear physics, Dover Pub., New York, 2010.
\bibitem{Bassi63}
P. Bassi {\it et al.}, Nuovo Cim. \textbf{28}, 1049 (1963).
\bibitem{King77}
C.H. King {\it et al.}, Phys. Rev. C \textbf{16}, 1712 (1977).
\bibitem{Illidis07}
C. Iliadis, Nuclear Physics of Stars, Wiley-VCH Verlag, Weinheim, 2007.
\bibitem{Slob82}
R.J. Slobodrian and H.E. Conzett, Z. Phys. A \textbf{308}, 15 (1982).
\bibitem{CA63}
Y. Cassagnou {\it et al.}, Nucl. Phys. \textbf{41}, 176 (1963).
\bibitem{DirectLi7p}
S. Engstler {\it et al.}, Z. Phys. A \textbf{342}, 471 (1992).
\bibitem{Bahcall69}
N.A. Bahcall and W.A. Fowler, Astrophys. J. \textbf{157}, 645 (1969).
\bibitem{Rolfs88}
C.E. Rolfs and W.S. Rodney, Cauldrons in the Cosmos, University of Chicago Press, 1988.
\bibitem{RCWFN}
A.R. Barnett {\it et al.}, Comp. Phys. Commun. \textbf{8}, 377 (1974).
\bibitem{Lax48}
M. Lax and H. Feshbach, J. Acoust. Soc. Am. \textbf{20}, 108 (1948).
\bibitem{r0}
A. Kamal, Nuclear Physics (Graduate Texts in Physics), Springer-Verlag Berlin and Heidelberg, 2014.
\bibitem{FCZ67}
W.A. Fowler, G.R. Caughlan and B.A. Zimmerman, Ann. Rev. Astron. Astrophys. \textbf{5}, 525 (1967).
\bibitem{Vogl64}
W.A. Fowler and J.L. Vogl, Lectures in Theoretical Physics VI, Univ. Clorado Press, 1964.
\bibitem{gale}
M. Gai and L. Weissman, Letter of Intent on SARAF facility (Soreq Nuclear Center, Israel), UConn-40870-00XX, May 5th, 2011.
\bibitem{cern}
M. Barbagallo and A. Musumarra {\it et al.}, CERN INTC meeting, June 25th, 2014.
\bibitem{TENDL}
TENDL-2014: TALYS-based evaluated nuclear data library,  ftp://ftp.nrg.eu/pub/www/talys/tendl2014/tendl2014.html.
\bibitem{Exp2}
T. Rauscher, EXP2RATE V2.1, http://nucastro.org/codes.html.
\bibitem{rau00}
T. Rauscher and F.-K. Thielemann, At. Data Nucl. Data Tables \textbf{75}, 1 (2000).
\bibitem{Hou10}
S.Q. Hou {\it et al.}, Chin. Phys. Lett. \textbf{27}, 082601 (2010).
\bibitem{WMAP9}
G. Hinshaw {\it et al.}, Astrophys. J. Suppl. \textbf{208}, 19 (2013).
\bibitem{Tau14}
K.A. Olive {\it et al.} (Particle Data Group), Chin. Phys. C \textbf{38}, 090001 (2014).
\bibitem{NvKind}
LEP Collaboration, Phys. Rep. \textbf{427}, 257 (2006).
\bibitem{Nacre99}
C. Angulo {\it et al.}, Nucl. Phys. A \textbf{656}, 3 (1999).
\bibitem{kusa13}
M. Kusakabe {\it et al.}, Phys. Rev. D \textbf{87}, 085045 (2013).
\bibitem{coc13}
A. Coc {\it et al.}, Phys. Rev. D \textbf{87}, 123530 (2013).
\end{thebibliography}
\end{document}